\documentclass[twocolumn,prb,floats,showpacs,superscriptaddress]{revtex4}
\usepackage{graphicx,epsfig}% Include figure files
\usepackage{times}
\usepackage{graphics,dcolumn,bm,float}
\usepackage{amssymb,amsmath,rotate,color}

\def\dblone{\hbox{$1\hskip -1.8pt\vrule depth 0pt height 1.6ex width 0.7pt \vrule depth 0pt height 0.3pt width 0.12em$}}
\begin{document}

\title{Optoelectronic properties of defective MoS$_2$ and WS$_2$ monolayers}
\author{Saboura Salehi}
\affiliation{Department of Physics, Payame Noor University, P.O.
Box 19395-3697 Tehran, Iran}
\author{Alireza Saffarzadeh}\email{asaffarz@sfu.ca}
\affiliation{Department of Physics, Payame Noor University, P.O.
Box 19395-3697 Tehran, Iran} \affiliation{Department of Physics,
Simon Fraser University, Burnaby, British Columbia, Canada V5A
1S6}
\date{\today}

\begin{abstract}
We theoretically explore the effect of metal and disulphur
vacancies on electronic and optical properties of MoS$_2$ and
WS$_2$ monolayers based on a Slater-Koster tight-binding model and
including the spin-orbit coupling. We show that the vacancy
defects create electronic flat bands by shifting the Fermi level
towards the valence band, indicating that both types of vacancies
may act as acceptor sites. The optical spectra of the pristine
monolayers show step-like features corresponding to the transition
from spin split valence band to the conduction band minimum,
whereas the defective monolayers exhibit additional peaks in their
spectra arising from induced midgap states in their band
structures. We find that Mo and W vacancies contribute mostly in
the low-energy optical spectrum, while the S$_2$ vacancies enhance
the optical conductivity mainly in the visible range of the
spectrum. This suggests that depending on the type of vacancy, the
atomic defects in MoS$_2$ and WS$_2$ monolayers may increase the
efficiency of solar cells used in photovoltaic systems.

\end{abstract}
%\pacs{71.55.Ak, 71.20.Mq, 31.15.aq}

\maketitle

\section{Introduction}
Two-dimensional (2D) transition metal dichalcogenides (TMDs) have
attracted increasing attention because of their unique physical
and chemical properties \cite{Radisavljevic,Braga}. These
materials with the general formula MX$_2$ have a layered structure
in the form of X-M-X where M is a transition metal from group IV,
V, or VI and X is a chalcogen. Due to their indirect-to-direct
bandgap and tunable electronic properties, they can be potentially
used for many electronic and optoelectronic applications. For
instance, MoS$_2$ and WS$_2$ monolayers with band gaps allowing
absorption in the visible region of the electromagnetic spectrum
can be utilized in field-effect transistors, solar cells, and
electroluminescent devices \cite{Wang,Ataca,Tsai,Splendiani}.

Due to the presence of spin-orbit coupling in monolayer of MX$_2$,
their valence band is split at the K-point of the Brillouin zone
and therefore, direct transitions between the maxima of the split
valence bands and the minimum of the conduction band can occur
\cite{Wang,Mak2010,Splendiani,Dhakal}. In this regard, recent
optical measurements of MoS$_2$ have shown two main peaks
corresponding to A and B exciton bands which are related to such a
spin splitting \cite{Mak2010,Splendiani}. In fact, the
experimental absorption spectra have focused mainly on the A and B
exciton transitions. Moreover, optical properties of pristine TMD
monolayers have also been theoretically studied by several groups
using first principles calculations
\cite{Qiu2013,Berkelbach,Gibertini,Amara}, Slater-Koster
tight-binding model \cite{Trolle,Guillen} and equation of motion
approach \cite{Chaves}.

On the other hand, structural defects, such as vacancies,
dislocations, and grain boundaries have been observed in pristine
TMD monolayers \cite{Qiu,Zhou,Van}. Application of these defects
to improve the optoelectronic properties of the monolayers is
highly desirable. Accordingly, atomic defects in TMD materials
have been studied both theoretically and experimentally
\cite{Yuan,Kunstmann,Srivastava,He1,Khan}.

Vacancy defects in TMD monolayers can be created by thermal
annealing and $\alpha$ particles \cite{Tongay} or electron and ion
beam irradiation \cite{Zhou,Ma1,Ma2}. Introduction of these
defects in the structures forms localized trap states in the
bandgap region which lead to light emission at energies lower than
the interband optical transition energy \cite{Tongay,Ma2}. In this
context, Khan et al. \cite{Khan} have recently studied optical
susceptibilities of single-layer TMDs in the presence of vacancy
defects using first-principles calculations. They showed that the
localized defect states induce sharp transitions in optical
susceptibility spectrum and a simple tight-binding model can
reveal the essential features of the localized defect states in
electronic band structures. The localized excitonic states related
to vacancy defects can tune the band gap \cite{Huang} and serve as
single-photon emitters \cite{Srivastava,He1}. Experimental
observations have shown that monosulphur vacancies are the most
abundant defects. However, their behavior as an acceptor
\cite{Komsa} or a donor
\cite{Suh,Radisavljevic,Liu2,Lu,Qiu,Salehi} is still under debate.
Indeed, charge transfer with substrate and also impurities such as
rhenium which is present in the environment can play a significant
role in the type of doping in TMD monolayers with S vacancies
\cite{Komsa}. Nevertheless, the Mo and W point defects can make
the system a p-type semiconductor \cite{Lu,Komsa}.

In this work, based on a parameterized tight-binding model which
includes the spin-orbit coupling and ignores the electron-hole
interaction, we investigate the influence of metal and disulphur
vacancy defects at two different concentrations on electronic and
optical properties of MoS$_2$ and WS$_2$ monolayers. Although
disulphur vacancies in TMD materials may lead to recombination of
surface atoms and cause some distortions in the monolayer
structures \cite{Zhou}, such effects are ignored in our theory. It
is shown that the atomic defects strongly affect the joint density
of states (JDOS) and optical properties of the pristine monolayers
by inducing mid-gap states in the band structure of these
materials. Moreover, the step-like features in the optical
spectrum of defect-free structures, induced by spin-orbit coupling
may change by introducing the vacancy defects. To do this, in
Section 2 we introduce our model and formalism for calculation of
electronic band structure, JDOS, and optical conductivity of the
pristine and defective monolayers. Numerical results and
discussion for both MoS$_2$ and WS$_2$ monolayers with different
types and concentrations of vacancy defects are presented in
Section 3. A brief conclusion is given in Section 4.

\begin{figure}
\centerline{\includegraphics[width=0.9\linewidth]{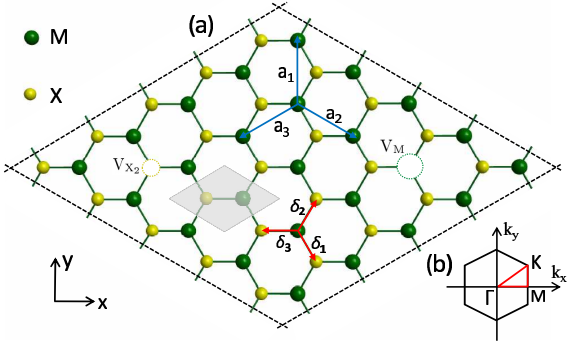}}
\caption{(Color online) (a) Top view of a 5$\times$5 single-layer
MX$_2$ supercell containing a single M or X$_2$ vacancy
($\mathrm{V_{M}}$ or $\mathrm{V_{X_2}}$). The nearest and the next
nearest neighbors are shown by vectors $\boldsymbol{\delta}_i$ and
${\bf a}_i$, respectively. The grey area shows the unit cell of
pristine MX$_2$ monolayer. (b) Hexagonal Brillouin zone of a
MX$_2$ monolayer with the red lines along which the band
structures are calculated.}
\end{figure}

\section{Model and formalism}
To study the electronic and optical properties of defective
MoS$_2$ and WS$_2$ monolayers we use a six-band tight-binding
approximation which is computationally inexpensive compared to
other approaches such as 11-band method \cite{Cappelluti,Roldan}
and non-orthogonal model with $sp^3d^5$ orbitals \cite{Zahid}. The
model is based on the three 4$d$-orbitals of the transition metal
atoms (Mo/W), $d_{xy}$, $d_{x^2-y^2}$, and $d_{3z2-r^2}$ with
dominant contribution in the valence and conduction bands of
MX$_2$ monolayers around the Fermi energy, and also the symmetric
(antisymmetric) combinations of 3$p$-orbitals of the top and
bottom chalcogen atoms S, i.e., $p_x$, $p_y$ $(p_z)$
\cite{Cappelluti,Roldan,Gomez,Guillen}. The basis vector for
electrons with spin $\sigma=\uparrow,\downarrow$ can be expressed
as
\begin{equation}\label{psi}
|\Phi_{\sigma}\rangle=(d_{3z^2-r^2,\sigma},d_{x^2-y^2,\sigma},d_{xy,\sigma},
p^+_{x,\sigma},p^+_{y,\sigma},p^-_{z,\sigma})^T
\end{equation}
where $p^\nu_i=(p^t_i+\nu p^b_i)/\sqrt{2}$ with $\nu=\pm$,
$i=x,y,z$, and $t$($b$) refers to the top (bottom) chalcogen atom.
In order to model a defective MX$_2$ monolayer with different
vacancy concentrations the structure is partitioned into $n\times
n$ supercells each containing $n^2$ unit cells where $n$ is an
integer number. Fig. 1(a) shows a 5$\times$5 supercell containing
X or M vacancies. An atomic vacancy can be modeled by removing one
X or M atom from each supercell without any relaxation of atomic
positions. Considering the basis set in Eq. (\ref{psi}), the
tight-binding Hamiltonian of the MX$_2$ monolayers can be written
as
\begin{equation}\label{H}
H=H_{SK}\otimes\dblone+H_{SO}\ ,
\end{equation}
with
\begin{equation}\label{HSK}
H_{SK}=\sum_{i,j}\sum_{\alpha,\beta}(\epsilon_{i\alpha}\delta_{ij}
\delta_{\alpha\beta}+t_{i\alpha,j\beta})b^\dag_{i\alpha}b_{j\beta}\
,
\end{equation}
\begin{equation}\label{HSO}
H_{SO}=\sum_{i}\sum_{\sigma,\sigma'}\frac{\lambda_i}
{2\hbar}\mathbf{L}_{i}\cdot\bm{\tau}_{\sigma\sigma'}\ ,
\end{equation}
where $H_{SK}$ and $H_{SO}$ represent the Slater-Koster
tight-binding Hamiltonian \cite{Slater} and the spin-orbit
coupling term, respectively, and $\dblone$ is the $2\times2$
identity matrix for spin of electron. Here,
$b^\dag_{i\alpha}(b_{i\alpha})$ is the creation (annihilation)
operator for an electron in the atomic orbital $\alpha$ at $i$-th
atom, $\varphi_{i\alpha}$, with on-site energy
$\epsilon_{i\alpha}$. The interatomic hopping parameters,
$t_{i\alpha,j\beta}=\langle\varphi_{i\alpha}|H_{SK}|\varphi_{j\beta}\rangle$,
between atomic orbitals $\varphi_{i\alpha}$ and $\varphi_{j\beta}$
depend on the Slater-Koster TB parameters and directional cosines
of the vector connecting the corresponding atoms. These parameters
are related to hopping processes between nearest neighbors M-M,
X-X, and M-X. See vectors $\mathbf{a}_i$ and
$\boldsymbol{\delta}_i$ ($i$=1, 2, and 3) in Fig. 1(a). The
hopping terms between next nearest neighbors are ignored in this
model. In Eq. \ref{HSO}, $\bm{\tau}$ are the Pauli spin matrices,
$\mathbf{L}_{i}$ is the atomic angular momentum operator, and
$\lambda_i$ is the intra-atomic spin-orbit coupling constant which
depends on the type of atom $i$. The spin-orbit coupling leads to
a strong spin-splitting in the valence-band maximum (VBM),
especially in WS$_2$ monolayers \cite{Cao}. Therefore, the
inclusion of this interaction which is necessary for obtaining the
correct band structures can affect optical spectrum of the TMD
monolayers by inducing new optical transitions.

The spin-dependent Hamiltonian in $\textbf{k}$-space for a
pristine MX$_2$ monolayer is given by
$H=\sum_{\mathbf{k},\sigma}H(\mathbf{k},\sigma)$ with
\begin{equation}\label{Hk}
H(\mathbf{k},\sigma)=\left(%
\begin{array}{cc}
H_{\mathrm{MM}}(\mathbf{k},\sigma) & H_{\mathrm{MX}}(\mathbf{k}) \\
H_{\mathrm{MX}}^\dag(\mathbf{k})   & H_{\mathrm{XX}}(\mathbf{k},\sigma) \\
\end{array}%
\right)\  ,
\end{equation}
\begin{equation}\label{HMMXX}
H_{\mathrm{MM(XX)}}(\mathbf{k},\sigma)=E_{\mathrm{M(X)}}(\sigma)
+2\sum_{i=1}^{3}T_i^{\mathrm{MM(XX)}}\cos(\mathbf{k}\cdot\mathbf{a}_i)\
,
\end{equation}
\begin{equation}\label{HMX}
H_{\mathrm{MX}}(\mathbf{k})=\sum_{i=1}^{3}T_i^{\mathrm{MX}}\exp(-i\mathbf{k}\cdot\boldsymbol{\delta}_i)\
,
\end{equation}
where $E_{\mathrm{M(X)}}$, $T_i^{\mathrm{MM(XX)}}$, and
$T_i^{\mathrm{MX}}$ are $3\times3$ $\mathbf{k}$-independent
matrices. These matrices depend on the Slater-Koster parameters
only and are given in Ref. [\onlinecite{Guillen}]. Note that
although the band structures of pristine MoS$_2$ and WS$_2$
presented in Ref. [\onlinecite{Guillen}] are correct, the given
Slater-Koster parameters therein cannot reproduce the correct band
structure of WS$_2$. Therefore, the parameters for MoS$_2$ and
WS$_2$ monolayers with a minor correction are given in Table. I.

To investigate the optical response of a defective TMD monolayer
to a linearly polarized light we study joint density of states
(JDOS) and real part of optical conductivity. The JDOS provides a
measure of the number of allowed optical transitions between the
occupied electronic states of the valence band and the unoccupied
electronic states of the conduction band separated by energy
$\hbar\omega$:

\begin{equation}\label{JDOS}
\mathrm{JDOS}(\omega)=\sum_{mn\sigma}\sum_{\mathbf{k}}
\delta(\varepsilon^{\sigma}_{mn}({\bf{k}})-\hbar\omega)\
,
\end{equation}
where $\omega$ is the light frequency and
$\varepsilon^{\sigma}_{mn}({\bf{k}})=\varepsilon^{\sigma}_{m}({\bf{k}})-\varepsilon^{\sigma}_{n}({\bf{k}})$
is the transition energy between occupied band $n$ and empty band
$m$. Therefore, the JDOS represents the number of states that can
undergo energy and ${\bf k}$-conserving (direct) transitions for
photon frequencies between $\omega$ and $\omega+d\omega$.

Moreover, in the electric-dipole approximation, the momentum
matrix element for an interband electronic transition from the
eigenstate $|n,\mathbf{k},\sigma\rangle$ in an occupied band to
the eigenstate $|m,\mathbf{k},\sigma\rangle$ in an empty band is
given by $p^{\sigma}_{x,mn}=\langle m,{\bf
k},\sigma|\hat{p}_x|n,{\bf k},\sigma\rangle$ which is assumed that
the electromagnetic field is along the $x$-direction. The momentum
operator is given by $\hat{\mathbf{p}}=(m/\hbar)\nabla
H(\mathbf{k},\sigma)$. Note that the matrix elements
$p^{\sigma}_{x,mn}$ determine whether the transition is allowed or
forbidden. A zero momentum-matrix element means a forbidden
transition. However, the transition is allowed when the symmetry
of $p^{\sigma}_{x,mn}$ spans the totally symmetric representation
of the point group to which the unit cell belongs. Therefore, the
real part of optical conductivity can be obtained from the Kubo
formula \cite{Yu}; i.e.,
\begin{equation}\label{Re}
\sigma_{xx}(\omega)=\frac{e^2}{4\pi
m^2\omega}\sum_{mn\sigma}\int_{1\mathrm{BZ}}
|p^{\sigma}_{x,mn}|^2\delta(\varepsilon^{\sigma}_{mn}({\bf{k}})-\hbar\omega)d^2\bf{k}\
,
\end{equation}
where $e$ is the electron charge and $m$ is the electron mass. The
two-dimensional integral is carried over the first Brillouin zone.

Note that the Dirac delta function in Eqs. (\ref{JDOS}) and
(\ref{Re}) is approximated by a Lorentzian function with a
broadening factor of 0.005 eV. In addition, we use a $500\times
500$ \textbf{k}-point mesh for accurate Brillouin zone
integrations.

\begin{table}[h]
\caption{The spin-orbit coupling (SOC) and Slater-Koster
tight-binding parameters of MoS$_2$ and WS$_2$ monolayers (taken
from Ref. [\onlinecite{Guillen}] with a minor correction in
$V_{pd\sigma}$ of W-S). All values are in units of eV.} \centering
\begin{tabular}{c c rr}
\hline\hline
&  &~~~~MoS$_2$~~&WS$_2$~~~~\\
\hline
&~~$\lambda_M$~~&~~0.086~~~&~~~0.271~~~\\[-1.2ex]
\raisebox{1.5ex}{SOC} & $~~\lambda_X~~$
&~~0.052~~~&~~~0.057~~~\\[0.05ex]
\hline
& $\Delta_0$ &~~-1.094~~~&~~~-1.155~~~\\[0.0ex]
& $\Delta_1$ &~~-0.050~~~&~~~-0.650~~~\\[-0.0ex]
Crystal Fields~~~&$\Delta_2$ &~~-1.511~~~&~~~-2.279~~~\\[-0.0ex]
& $\Delta_p$ &~~-3.559~~~&~~~-3.864~~~\\[-0.0ex]
& $\Delta_z$ &~~-6.886~~~&~~~-7.327~~~\\[0.05ex]
\hline
& $V_{pd\sigma}$ &~~3.689~~~&~~~4.911~~~\\[-1ex]
\raisebox{1.5ex}{M-S} & $V_{pd\pi}$
&~~-1.241~~~&~~~-1.220~~~\\[0.05 ex]
\hline
& $V_{dd\sigma}$ &~~-0.895~~~&~~~-1.328~~~\\[0.0ex]
M-M & $V_{dd\pi}$ &~~0.252~~~&~~~0.121~~~\\[0.0ex]
& $V_{dd\delta}$ &~~0.228~~~&~~~0.442~~~\\[0.05ex]
\hline
&$V_{pp\sigma}$ &~~1.225~~~&~~~1.178~~~\\[-1.2ex]
\raisebox{1.5ex}{S-S} & $V_{pp\pi}$
&~~-0.467~~~&~~~-0.273~~~\\[0.05ex]
\hline\hline
\end{tabular}
\label{tab:1}
\end{table}

\begin{figure}
\centerline{\includegraphics[width=0.95\linewidth]{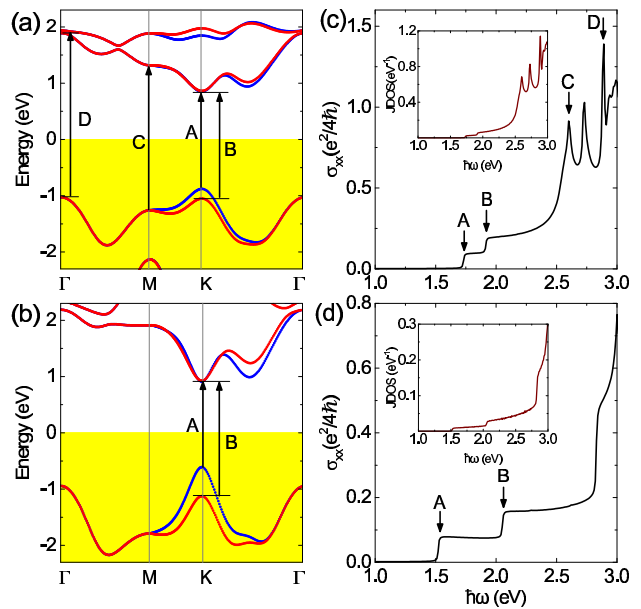}}
\caption{(Color online) (a)((b)) Calculated band structure with
projection of spin operator and (c)((d)) the real part of optical
conductivity of pristine MoS$_2$ (WS$_2$) monolayer. The red and
blue colors in the band structures indicate the spin-up and
spin-down states, respectively. The insets show the corresponding
JDOS. The intersection of white and yellow regions in (a) and (b)
shows the Fermi energy. The arrows A, B, C, and D denote some
important interband optical transitions.}
\end{figure}

\section{results and discussion}
To study the influence of atomic vacancy defects on the electronic
and the optical properties of MoS$_2$ and WS$_2$ monolayers we
first discuss the band structure, JDOS, and optical conductivity
of the pristine monolayers. Figs. 2(a) and (b) show the band
structures with the projection of spin operator in the energy
range of $|E|\leq$ 2.3 eV for MoS$_2$ and WS$_2$ monolayers,
respectively. The splitting of the valence band at the K point in
WS$_2$ is more than two times larger than that in MoS$_2$, which
makes WS$_2$ monolayers a potential candidate for spin- and
valley-based electronic devices \cite{Xiao}. Note that the order
of spin bands in the band structures along $\Gamma$-K$^\prime$-M
(not shown here) will flip compared to those along the
$\Gamma$-K-M lines \cite{Salehi}. This feature leads to a
valley-selective optical absorption using a circularly polarized
light which may cause optically induced valley and spin Hall
effects \cite{Sie}. The difference in the valence band
spin-splitting between MoS$_2$ and WS$_2$ monolayers is directly
reflected in their JDOS and optical spectra, as shown in Figs.
2(c) and 2(d). The interband transitions from the valence bands to
the conduction bands can produce plateaus (shoulders) at lower
energies and a series of sharp peaks at higher energies in the
optical spectra which are attributed to the van Hove singularities
in the JDOS, shown in the insets of Fig. 2. The arrows, marked by
A, B, C, and D in the band structures of the two compounds, are
related to some of these transitions. Comparison of the van Hove
singularities in the JDOS with the plateaus and sharp peaks in the
optical conductivity reveals that the related optical transitions
of the pristine monolayers in the given energy range are all
allowed. The interband transitions A and B correspond to the
direct excitonic transitions which have been observed in the
photoluminescence measurements \cite{Wang,Mak1}. Note that in
order to qualitatively treat the excitonic effects in 2D materials
such as monolayers MoS$_2$ and WS$_2$, the effective Coulomb
interaction between electrons and holes should be included in the
theory \cite{Trolle}.

It is worth mentioning that although the momentum matrix elements
$p^{\sigma}_{x,mn}$ and $p^{\sigma}_{y,mn}$ have different
\textbf{k}-dependencies, the \textbf{k} integral over the whole
Brillouin zone in the Kubo formula \cite{Yu} gives the same
optical conductivities $\sigma_{xx}(\omega)$ and
$\sigma_{yy}(\omega)$, indicating that the optical spectra of
pristine MX$_2$ monolayers are isotropic. Such a feature has also
been reported for graphene at low and high frequencies
\cite{Nguyen}.
\begin{figure}
\center\includegraphics[width=0.95\linewidth]{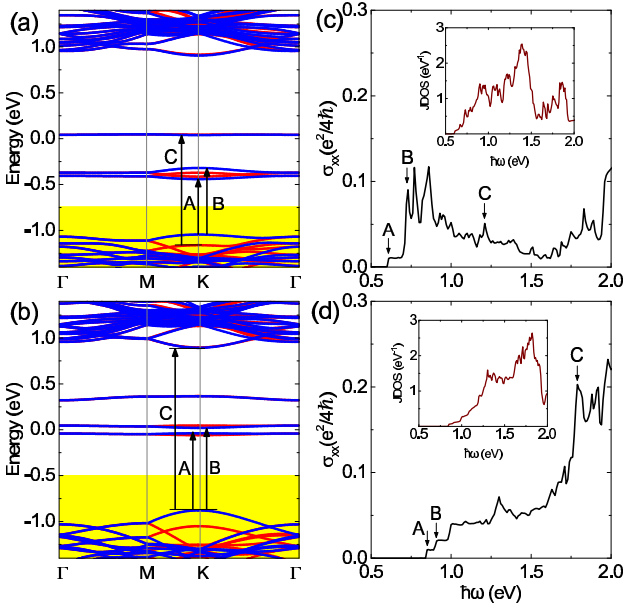}
\caption{Calculated band structures, optical conductivities, and
joint densities of states (insets) of MoS$_2$ monolayer with
5$\times$5 supercells for (a,c) $\mathrm{V_{Mo}}$ and (b,d)
$\mathrm{V_{S_2}}$. The arrows A, B, and C denote some important
interband optical transitions. The intersection of white and
yellow regions in (a) and (b) shows the Fermi energy.}
\end{figure}

Now, we investigate the influence of Mo, W, and S$_2$ vacancies on
the optical properties of the layered structures. To control the
vacancy concentration, we use the supercell approach in which a
perfect  monolayer is described as an infinite periodic array of
$n\times n$ supercells \cite{Salehi}. Then a vacancy defect is
created by removing one atom from each supercell without any
change in the symmetry of the lattice (see Fig. 1(a)). Note that
in the supercell approach the corresponding Brillouin zone shrinks
in size and the bands get folded. We refer to a single Mo(W)
vacancy as $\mathrm{V_{Mo(W)}}$, whereas vacancies of two sulphur
atoms (one at the top layer and the other at the bottom layer of
MX$_2$) attached to the same Mo/W atom are denoted as
$\mathrm{V_{S_2}}$ (see Fig. 1).

Figs. 3(a) and (b) show the band structures of MoS$_2$ in
5$\times$5 supercells with $\mathrm{V_{Mo}}$ and
$\mathrm{V_{S_2}}$, respectively. For this case, the concentration
of $\mathrm{V_{Mo}}$ is $\frac{1}{75}$, while it is $\frac{2}{75}$
for $\mathrm{V_{S_2}}$, corresponding to defect densities $\sim
4.6\times10^{13}$ cm$^{-2}$ and $9.2\times10^{13}$ cm$^{-2}$,
respectively. The vacancy states which are mainly localized around
atomic defects \cite{Yuan} induce several narrow (flat) bands for
each spin subband around zero energy in the band structures
\cite{Note}. Although the midgap states are almost degenerate, the
spin splitting in the valence bands makes some of the interband
transitions spin dependent (e.g., see transition C in Fig. 3(a)).
The tendency of defect states in Fig. 3(a) towards the valence
band predicts the $p$-type semiconducting behavior by introducing
Mo vacancies in this structure. For the case of
$\mathrm{V_{S_2}}$, however, there is a tendency towards the
primary conduction band (see Fig. 3(b)). We should mention that
the position of Fermi energy, shown in the figures, is determined
by counting the number of electrons that the atoms in the
supercell provide. These electrons fill up the lowest spin
dependent energy bands and hence, the Fermi level lies between the
highest occupied band and the lowest unoccupied band. In other
words, since the energy bands are spin dependent, all the bands
below the Fermi energy are singly occupied.

Optical conductivity and JDOS of MoS$_2$ monolayer with
$\mathrm{V_{Mo}}$ and $\mathrm{V_{S_2}}$ are depicted in Figs.
3(c) and (d), respectively. Some interband transitions in the band
structures, marked by A, B, and C are shown in the optical
spectra. The optical transitions for the structure with
$\mathrm{V_{Mo}}$ can occur at photon energies $\hbar\omega\geq
0.6$eV, whereas, these transitions for the monolayer with
$\mathrm{V_{S_2}}$ start at $\hbar\omega=0.84$eV. Due to more flat
bands in the band gap, the sulphur vacancies induce more plateaus
compared to $\mathrm{V_{Mo}}$. Moreover, the joint densities of
states show different features compared to the optical spectra,
indicating that there are considerable forbidden transitions in
the defective MoS$_2$ spectrum as discussed above.
\begin{figure}
\center\includegraphics[width=0.95\linewidth]{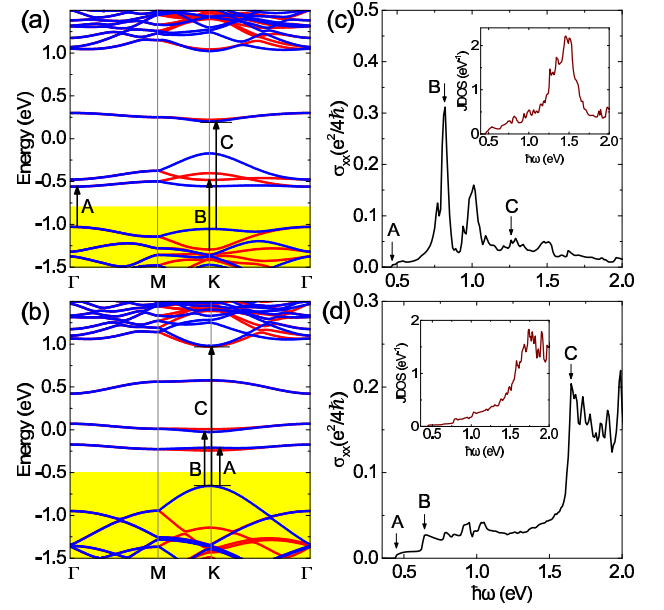}
\caption{Calculated band structures, optical conductivities, and
joint densities of states (insets) of WS$_2$ monolayer with
5$\times$5 supercells for (a,c) $\mathrm{V_{W}}$ and (b,d)
$\mathrm{V_{S_2}}$. The arrows A, B, and C denote some important
interband optical transitions. The intersection of white and
yellow regions in (a) and (b) shows the Fermi energy.}
\end{figure}

The band structures of WS$_2$ monolayers with $\mathrm{V_{W}}$ and
$\mathrm{V_{S_2}}$ are shown in Figs. 4(a) and (b). We see that
the difference between the valence band maximum and the lowest
defect state in energy is almost the same for both vacancies.
Although there is a tendency towards valence band for the
$\mathrm{V_{W}}$-related midgap states, such a tendency is weaker
for the $\mathrm{V_{Mo}}$-related states in MoS$_2$ monolayer. On
the contrary, the energy difference between the conduction band
minimum and the highest $\mathrm{V_{S_2}}$-related state is almost
the same as that between the valence band maximum and the lowest
midgap state. Optical conductivity and the joint densities of
states are shown in Figs. 4(c) and (d). As can be seen, the energy
of first interband transition (marked A) is almost the same for
$\mathrm{V_{W}}$ and $\mathrm{V_{S_2}}$ in WS$_2$ structure. Due
to the $\mathrm{V_{S_2}}$-related flat bands in the band gap, the
optical conductivity spectrum exhibits mainly a step-like behavior
for photon energies $\hbar\omega\leq$ 1.2 eV, compared to that for
the monolayer with $\mathrm{V_{W}}$. This feature can also be seen
in the JDOS of the structure containing $\mathrm{V_{S_2}}$. Again,
there are some forbidden optical transitions in the defective
structures with whether W or S$_2$ vacancies, due to the
remarkable differences between joint densities of states and the
optical spectra.

The optical conductivity can sharply reach a high value at
low-energy spectrum when W vacancies are introduced. In this case,
an optical transition, marked B in Figs. 4(a) and (c), from
valence band to midgap states occurs. However, in the case of
S$_2$ vacancies the conductivity increases sharply when a
transition from valence band to conduction band happens (see the
transition C in Fig. 4(b) and(d)). Qualitatively, similar features
can also be seen in the optical spectra of the defective MoS$_2$
structure (see Fig. 3). This suggests that the metal vacancies
contribute mainly in the low-energy optical spectrum, while the
disulphur vacancies contribute mainly in the transitions with
higher energies. Therefore, depending on the type of vacancy, the
atomic defects in MoS$_2$ and WS$_2$ monolayers as sunlight
absorbers may increase the efficiency of solar cells used in
photovoltaic systems \cite{Bernardi}.

\begin{figure}
\center\includegraphics[width=1.0\linewidth]{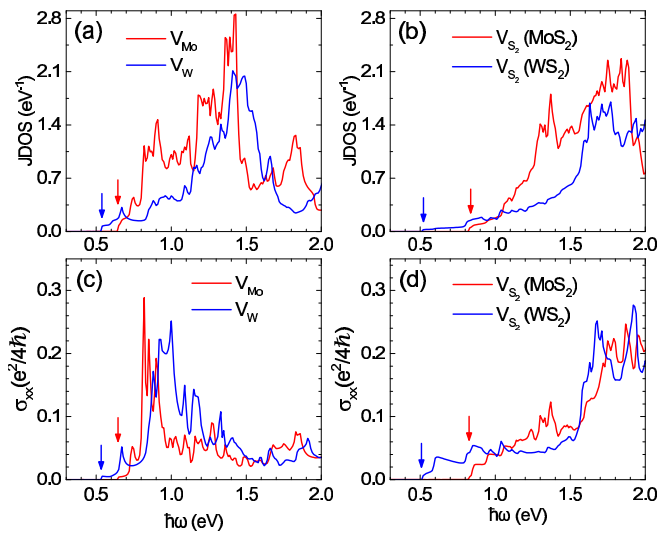} \caption{
(a,b) Joint densities of states and (c,d) optical conductivities
of defective monolayers MoS$_2$ (red lines) and WS$_2$ (blue
lines) with 4$\times$4 supercells. The arrows indicate the
energies at which the optical transitions begin.}
\end{figure}

We have also performed the tight-binding calculations on a smaller
supercell, i.e., $4\times4$, which corresponds to the higher
vacancy densities $\sim 7.2\times 10^{13}$ cm$^{-2}$
($\mathrm{V_{Mo/W}}$) and $14.4\times 10^{13}$ cm$^{-2}$
($\mathrm{V_{S_2}}$). The joint densities and optical conductivity
spectra for the monolayers with Mo, W, and S$_2$ vacancies are
depicted in Fig. 5. At higher defect densities the hybridization
of vacancy electronic states increases due to the reduction of
spatial separation between the vacant sites. As a result, the
energy of midgap states are shifted compared to that at low defect
concentration. To see this feature one can compare the energies at
which the first optical transition (or the first van Hove
singularity in the joint densities of states) occur in the
defective structures with $5\times 5$ and $4\times 4$ supercells.
In the structures with $4\times 4$ supercells, the energies of
first optical transitions, marked by red and blue arrows in Fig.
5, increase by 43 meV and 61 meV for Mo and W vacancies,
respectively, compared to those in Figs. 3 and 4. On the other
hand, the first transition for WS$_2$ structure with S$_2$
vacancies is 65 meV higher than that in Fig. 4, while it occurs 15
meV lower than that in Fig. 3. Therefore, with increasing the
density of metal vacancy in both structures or the density of
S$_2$ vacancy in WS$_2$ structure, the defect states shift towards
the conduction band, indicating that the tendency of $p$-type
semiconducting for these types of vacancies decreases with
increasing the defect concentration. The position and the strength
of optical peaks in Figs. 5(c) and (d) confirm our above
discussion that the metal vacancies considerably increase the
optical conductivity when the energy of photons is less than the
bandgap of the host structures, while the S$_2$ vacancies enhance
the optical conductivity mainly in the visible range of the
electromagnetic spectrum. Since the vacancy defects are
unavoidable during the synthesis of MX$_2$ monolayers, these
results are of great practical importance.

It is worth mentioning that the metal and disulphur vacancies do
not induce any magnetic moments in MoS$_2$ and WS$_2$ monolayers
\cite{Lu,Tao}. By applying a tensile strain \cite{Tao} or
substitutional doping \cite{Lu}, however, one can change the
magnetic properties of these defective structures. Moreover,
although the localized midgap states can act as scattering centers
and strongly affect electron transport through the material
\cite{Asl}, these defect states are responsible for an increase in
photoluminescence intensity \cite{Tongay}. In other words, the
defective structures can emit stronger light and are promising for
optoelectronic device applications such as light emitting diodes.

\section{conclusion}

Based on Slater-Koster six-band tight-binding model and including
the spin-orbit coupling, we have studied the effect of Mo, W, and
disulphur vacancies on electronic and optical properties of
MoS$_2$ and WS$_2$ monolayers. It is shown that the defect states
induce flat bands in the bandgap and shift the Fermi level towards
the valence band, suggesting that these vacancies may act as
acceptor sites. Moreover, the localized midgap states have the
potential to activate new optical transitions with energies less
than the bandgap of pristine structures. We found that although
the electronic behavior of MoS$_2$ and WS$_2$ monolayers in the
presence of both types of metal and disulphur vacancies are almost
the same, contributions of the metal and disulphur vacancies in
the optical conductivity occur in different ranges of incident
photon energy. The disulphur defects activate the optical spectrum
in the visible range, while the Mo and W vacancies activate the
conductivity mostly at low energies.

Therefore, our findings are able to reveal some electronic and
optical features of the defective MoS$_2$ and WS$_2$ monolayers
using a six-band tight-binding model which is less computationally
demanding than the traditional \textit{ab initio} methods.

\section*{Acknowledgement}
This work is partially supported by Iran Science Elites Federation
(11/66332).


\begin{thebibliography}{99}
\bibitem{Note}
Note that by taking a larger supercell such as $12\times12$, the
vacancy density $\sim 10^{12}$ cm$^{-2}$ will be more in line with
experimental values. In such a density, the distance between two
vacancies becomes considerably large and the interaction between
defects will be negligible. Therefore, one can expect that the
midgap states become completely dispersionless due to the strong
localization of defect states \cite{Salehi}.
\end{thebibliography}
\end{document}